\documentclass[prb,amsmath,amssymb,twocolumn,floatfix]{revtex4-1}
\usepackage{graphicx}% Include figure files
\usepackage{dcolumn}% Align table columns on decimal point
\usepackage{bm}% bold math
\usepackage{subfigure}
\usepackage{amsmath}
\usepackage{amssymb}
\usepackage{hyperref}
\usepackage{times}
\usepackage{xcolor}

\newcommand{\lan}{\langle}
\newcommand{\ran}{\rangle}

\newcommand{\ket}[1]{|#1 \rangle}

\newcommand{\mr}[1]{\mathrm{#1}}

\newcommand{\mcal}[1]{\mathcal{#1}}

\begin{document}
\title{Finite-Size Scaling Analysis of the Planck's Quantum-Driven Integer Quantum Hall Transition in Spin-$1/2$ Kicked Rotor Model}
\author{Jia-Long Zhang}
\affiliation{Kavli Institute for Theoretical Sciences and CAS Center for Excellence in Topological Quantum Computation, University of Chinese Academy of Sciences, Beijing 100190, China}
\author{Long Zhang}
\affiliation{Kavli Institute for Theoretical Sciences and CAS Center for Excellence in Topological Quantum Computation, University of Chinese Academy of Sciences, Beijing 100190, China}
\author{Fu-Chun Zhang}
\affiliation{Kavli Institute for Theoretical Sciences and CAS Center for Excellence in Topological Quantum Computation, University of Chinese Academy of Sciences, Beijing 100190, China}

\date{\today}

\begin{abstract}
The quantum kicked rotor (QKR) model is a prototypical system in the research of quantum chaos. In a spin-$1/2$ QKR, tuning the effective Planck parameter realizes a series of transitions between dynamical localization phases, which closely resembles the integer quantum Hall (IQH) effect and the plateau transitions. In this work, we devise and apply the finite-size scaling analysis to the transitions in the spin-$1/2$ QKR model. We obtain an estimate of the critical exponent at the transition point, $\nu=2.62(9)$, which is consistent with the IQH plateau transition universality class. We also give a precise estimate of the universal diffusion rate at the metallic critical state, $\sigma^{*}=0.3253(12)$.
\end{abstract}

\maketitle

\section{Introduction}

The kicked rotor model describes a particle moving on a circle and is kicked periodically by a space-dependent potential term. It is a prototypical model in the research of both classical and quantum chaos \cite{Chirikov:79, Izrailev:90, Casati:79}. For a sufficiently large kicking strength, the energy of a classical kicked rotor grows linearly with time, $E(t)\propto t$, as a result of the Brownian motion in the momentum space. However, such linear diffusive motion is suppressed in the long-time limit for a quantum kicked rotor (QKR) due to the destructive quantum interference, leading to a dynamical localization in the momentum space \cite{Fishman:82}. The kicked rotor model has been generalized to higher dimensions and spinful particles \cite{Shepelyansky:83, Casati:89, Scharf:89, Thaha:94, Masovic:94, Ossipov:04, Bardarson:05, Bardarson:07}.

A particular interesting discovery is the Planck's quantum-driven integer quantum Hall (IQH) effect in a spin-$1/2$ QKR model \cite{Chen:14, Tian:16}, which establishes a surprising bridge between chaotic systems characterized by the sensitivity to initial conditions, and the topologically robust IQH effect with a quantized Chern number. It is analytically shown that, by tuning the effective Planck's quantum $h_{e}$, the model defined in Eq. (\ref{eq:Ham}) exhibits an infinite number of ``Hall plateau'' transitions between the dynamically localized insulating phases. Each insulating phase is characterized by an integer $\sigma_{H}$ analogous to the quantized Hall conductance of IQH plateaus. The critical metallic states at the transition points are predicted to possess a universal ``longitudinal conductance'' $\sigma^{*}=\lim_{t\rightarrow\infty}E(t)/t$, and belong to the universality class of the IQH plateau transitions. The emergence of these transitions has been observed in numerical simulations \cite{Chen:14, Tian:16}. However, precise calculations of the critical exponents at the ``plateau transitions'' have not been achieved, thus leaving the universality class of the transitions not fully confirmed.

In this work, we apply the finite-size scaling analysis to the plateau transitions in the spin-$1/2$ QKR model. With extensive numerical simulations near the critical point and finite-size scaling analysis, we obtain an estimate of the critical exponent at the critical point $\nu=2.62(9)$, which is consistent with the IQH plateau transition universality class. We also give a precise estimate of the universal diffusion rate at the metallic critical state, $\sigma^{*}=0.3253(12)$.

The spin-$1/2$ QKR model is introduced in Sec. \ref{sec:model}, which can be obtained from a spin-$1/2$ kicked rotor in two dimensions (2D) by the dimension reduction technique. The numerical simulations and the finite-size scaling analysis are devised and applied to the QKR model in Sec. \ref{sec:numerics}, followed by a brief summary in Sec. \ref{sec:concl}.

\section{The Model}
\label{sec:model}

We shall study the spin-$1/2$ QKR model introduced in Refs. \onlinecite{Dahlhaus:11}. A spin-$1/2$ particle moves on a circle of unit radius and is kicked periodically by a potential field, whose strength depends on the position the particle. Denote the two-component spinor wavefunction by $\Psi_{t}$. Its dynamics is governed by the time-dependent Schr\"odinger equation,
\begin{equation}
ih_{e} \partial_{t} \Psi_{t} = H(t) \Psi_{t},
\end{equation}
in which the Hamiltonian is given by
\begin{equation}
H(t) = H_0(p_1,p_2)+V(\theta_1,\theta_2) \sum_{s\in \mathbb{Z}} \delta(t-s), \label{eq:Ham}
\end{equation}
where $\theta_1$ (modulo $2\pi$) is the angular position of the particle, and $p_{1}=-ih_{e}\partial_{\theta_{1}}$ is the conjugate momentum operator. While the model is defined in two space dimensions, we shall show in Sec. \ref{sec:red} that it can be simulated effectively in 1D with  the technique of dimension reduction. The canonical commutation relation is given by $[\theta_{i},p_{j}]=ih_e\delta_{ij}$. The effective Planck's constant $h_e$ is a tuning parameter in the model.

The generic form of the potential energy term is given by $V=V_i(\theta_1, \theta_2)\sigma^i$, where $\sigma^i$ ($i=1,2,3$) are the Pauli matrices. The Einstein summation convention is used. The potential energy term couples the spin and the angular position of the particle,
\begin{equation}
V(\theta_1,\theta_2)= (2 \arctan 2d/d)\boldsymbol{d}\cdot\boldsymbol{\sigma},
\label{eq:qwz_V}
\end{equation}
with the vector $\boldsymbol{d}$ given by
\begin{equation}
\boldsymbol{d} = (\sin \theta_1, \sin \theta_2, 0.8(\mu - \cos \theta_1 - \cos \theta_2)).
\end{equation}
This potential term of the spin-$1/2$ QKR model was first introduced in Ref. \onlinecite{Dahlhaus:11}, which was inspired by the Qi-Wu-Zhang model of quantum anomalous Hall effect \cite{Qi:06}. A series of phase transitions driven by the effective Planck parameter $h_e$ was found in Refs. \onlinecite{Chen:14, Tian:16}, which resembles the IQH plateau transitions in various significant aspects. In this work, we shall focus on the latter case and fix $\mu=1$ in the rest of this work.

\subsection{Floquet operator}

The nature of the long-time dynamics of the QKR can be obtained by inspecting the time-evolving state at integer time $t$. Given an initial state $\ket{\Psi_{0}}$ at $t_{0}=0$, the state at time $t$ can be obtained by applying the Floquet operator $t$ times on $|\Psi_0\rangle$, $|\Psi_t\rangle=\mathcal{F}^t |\Psi_0\rangle$, in which the Floquet operator $\mcal{F}$ is the time-evolution operator in one kicking period,
\begin{equation}
\mathcal{F} = e^{-i V(\boldsymbol{\theta})/h_e}e ^ {-i H_0(\boldsymbol{p}) /h_e}.
\end{equation}
In the angular position representation, $\boldsymbol{p}=-ih_e\partial_{\boldsymbol{\theta}}$. The Hamiltonian in Eq. (\ref{eq:Ham}) is $2\pi$-periodic in $\boldsymbol{\theta}$, thus the eigenstates of the Floquet operator $\mcal{F}$ can be decomposed in the following form due to the Floquet-Bloch theorem,
\begin{equation}
\Psi_{\boldsymbol{q}}(\boldsymbol{\theta}) = e^{i\boldsymbol{q}\cdot\boldsymbol{\theta}}u(\boldsymbol{\theta}),
\end{equation}
where $\boldsymbol{q}=(q_1,q_2)$ with the constants $q_{1,2} \in (0,1)$, and $u(\boldsymbol{\theta)}$ is a $2\pi$-periodic function of the angle variables $\boldsymbol{\theta}$. Therefore, for these eigenstates, $H_0(-ih_e\partial_{\boldsymbol{\theta}})$ can be replaced by $H_0(-ih_e\partial_{\boldsymbol{\theta}}+h_e\boldsymbol{q})$ acting on $u(\boldsymbol{\theta})$. The corresponding Floquet operator reads
\begin{equation}
\mathcal{F}_{\boldsymbol{q}} = e ^ {-i V(\boldsymbol{\theta})/h_e}e ^ {-i H_0(-ih_e\partial_{\boldsymbol{\theta}}+h_e\boldsymbol{q}) /h_e}.
\end{equation}

\subsection{Mapping to the Anderson model}

The QKR model can be mapped to the Anderson model of a particle moving in a quasi-disordered system, signifying the link between quantum chaos and Anderson localization \cite{Fishman:82, Grempel:84}. Let us first omit the spin degree of freedom and define the eigenstate of the Floquet operator by
\begin{equation}
\mathcal{F}_{q}|a_{+}\rangle = e^{-i\epsilon} |a_{+} \rangle,
\end{equation}
in which $\epsilon$ is called the quasi-energy. Here $|a_{+}\rangle$ is the eigenstate of the Floquet operator immediately after the kick. Define
\begin{equation}
|a_{-}\rangle = e^{iV} |a_{+}\rangle =e^{i\epsilon-iH_0}|a_{+}\rangle, \label{eq:am}
\end{equation}
which is the eigenstate before the kick, then $\ket{a_{\pm}}$ satisfy
\begin{equation}
|a_{+}\rangle = e^{-iV} |a_{-}\rangle \equiv \frac{1-iW}{1+iW} |a_{-}\rangle. \label{eq:ap}
\end{equation}
Define $|u\rangle = \frac{1}{2}(|a_{+}\rangle+|a_{-}\rangle)$, then we have
\begin{align}
|a_{+}\rangle &= (1-iW)|u\rangle, \\
|a_{-}\rangle &= (1+iW)|u\rangle.
\end{align}
Substituting into Eq. (\ref{eq:am}), we find $\ket{u}$ satisfies the following secular equation,
\begin{equation}
W|u\rangle = \tan \left( \frac{\epsilon-H_0}{2} \right) |u\rangle.
\end{equation}
In the momentum space with a basis $\{|\boldsymbol{n}\rangle\}$, where $\boldsymbol{p}|\boldsymbol{n}\rangle\ = h_e \boldsymbol{n}|\boldsymbol{n}\rangle$, and with the spin indices recovered, we arrived at
\begin{equation}
\sum_{\boldsymbol{n}s^\prime}W_{\boldsymbol{n}}^{ss^{\prime}} u_{\boldsymbol{n}+\boldsymbol{m}}^{s^\prime} + \tan \left(  \frac{H_0(\boldsymbol{m})-\epsilon}{2}  \right) u_{\boldsymbol{m}}^s=0,
\label{eq:ands}
\end{equation}
where $W_{\boldsymbol{n}-\boldsymbol{m}}^{ss^{\prime}}=\langle \boldsymbol{m},s| W|\boldsymbol{n}, s^\prime\rangle$, and $u_{\boldsymbol{m}}^s=\langle \boldsymbol{m},s|u, s\rangle$. This is an Anderson model in two dimensions, in which the kinetic term $H_{0}(\boldsymbol{m})$ in the QKR model plays the role of a quasi-disordered potential.

From Eq. (\ref{eq:ap}), the hopping matrix $W$ in the Anderson model is given by
\begin{equation}
W=i\frac{1-e^{iV}}{1+e^{iV}}=\tan(V/2).
\end{equation}
$W$ is diagonal in the angular position representation. Given the form of $V$ in \ref{eq:qwz_V}, we find
\begin{equation}
W=2\boldsymbol{d}\cdot \boldsymbol{\sigma}.
\end{equation}

\subsection{Dimension Reduction}
\label{sec:red}

The 2D QKR model can be effectively reduced to 1D by choosing an incommensurate driving frequency in the second dimension \cite{Shepelyansky:83, Casati:89, Borgonovi:97}. Consider the following separable kinetic energy term,
\begin{equation}
H_0(p_1, p_2) = H_0(p_1)+\omega p_2,
\end{equation} 
treat the second term as a ``non-interacting'' Hamiltonian, $\mathcal{H}_{0}=\omega p_2$, and the rest part as the ``interactions'', $\mathcal{H}_\text{int}=H_0(p_1)+V(\boldsymbol{\theta})\sum_{s\in \mathbb{Z}} \delta(t-s)$, and then transform into the interaction picture,
\begin{equation}
\Psi_{I}  = e^{i\mathcal{H}_0t/h_e}\Psi =e^{i\omega p_2 t/h_e}\Psi,
\end{equation}
and the transformed Hamiltonian is given by
\begin{equation}
\begin{split}
H_{I} &=  e^{i\mathcal{H}_0t/h_e}\mathcal{H}_\text{int} e^{-i\mathcal{H}_0t/h_e} \\
&=H_0(p_1) + V(\theta_1, \theta_2+\omega t)\sum_{s\in \mathbb{Z}} \delta(t-s),
\end{split}
\end{equation}
where the translation relation $e^{i\omega p_2 t/h_e} V(\theta_2) e^{-i\omega p_2 t/h_e} =V(\theta_2+\omega t)$ is used. The Schr\"odinger equation in the interaction picture reads
\begin{equation}
ih_e \partial_t \Psi_I  = H_I \Psi_I,
\end{equation}
in which the Hamiltonian is given by
\begin{equation}
H_I = H_0(p_1)+V(\theta_1,\theta_2+\omega t) \sum_{s\in \mathbb{Z}} \delta(t-s).
\end{equation}
This is a 1D model, which dramatically simplifies the following numerical calculations. The corresponding Floquet operator is given by
\begin{equation}
\mathcal{F}_{q} = e ^ {-i V(\theta_1,\omega t +\alpha)/h_e}e ^ {-i H_0(n_{1}+q)/h_e},
\end{equation}
where $n_1=p_1/h_e$. In the following numerical simulations, we adopt the kinetic term 
\begin{equation}
H_{0}(p_{1})=p_{1}^{2},
\end{equation}
and $\omega=\frac{2\pi}{\sqrt{5}}$, which is incommensurate with the driving frequency in first dimension. This guarantees that the disorder potential produced by the kinetic term $H_{0}(p_1, p_2)$ is sufficiently quasi-random to induce dynamical localization in the equivalent Anderson model.

\section{Numerical simulations}
\label{sec:numerics}

We work in the momentum representation in numerical simulations. The Hilbert space is truncated to be $2N$-dimensional such that the momentum index $n \in [-N,N-1]$. Two types of initial states are considered in our simulations. The first one is of the $\delta$-function form,
\begin{equation}
\Psi_0(n)=\lan n| \Psi_{0} \ran =  \delta_{n,0} 
\begin{pmatrix}
e^{-i\varphi/2} \cos(\phi/2) \\
e^{i\varphi/2} \sin(\phi/2)
\end{pmatrix},
\end{equation}
while the second is a Gaussian wavepacket given by
\begin{equation}
\Psi_0(n) \propto  e ^{-\frac{(n-n_0)^2}{2\sigma^2}} 
\begin{pmatrix}
e^{-i\varphi_n/2} \cos(\phi_n/2) \\
e^{i\varphi_n/2} \sin(\phi_n/2)
\end{pmatrix}
\end{equation}
up to a normalization factor. We set $n_0=0$ and $\sigma=1$ in our simulations.

The diffusion rate is defined by
\begin{equation}
D(t)=\frac{\Delta^2(t)}{t},
\end{equation}
where $\Delta^2(t)=\frac{1}{2}\overline{\lan \Psi_t|\hat{n}^2|\Psi_t\ran}$, and the rotor energy $E(t)=h_{e}^{2}\Delta^2(t)$. $\overline{\lan \cdots \ran}$ is the ensemble average over uniformly distributed $\alpha\in[0,2\pi)$ and $q\in[0,1)$, and the angle variables of the initial state $\varphi$ and $\phi$ (or $\varphi_{n}$ and $\phi_{n}$) uniformly distributed on the Bloch sphere. We find that both types of initial states give rise to the almost same diffusion rate after the ensemble average. 

\begin{figure}[tb]
\includegraphics[width=\columnwidth]{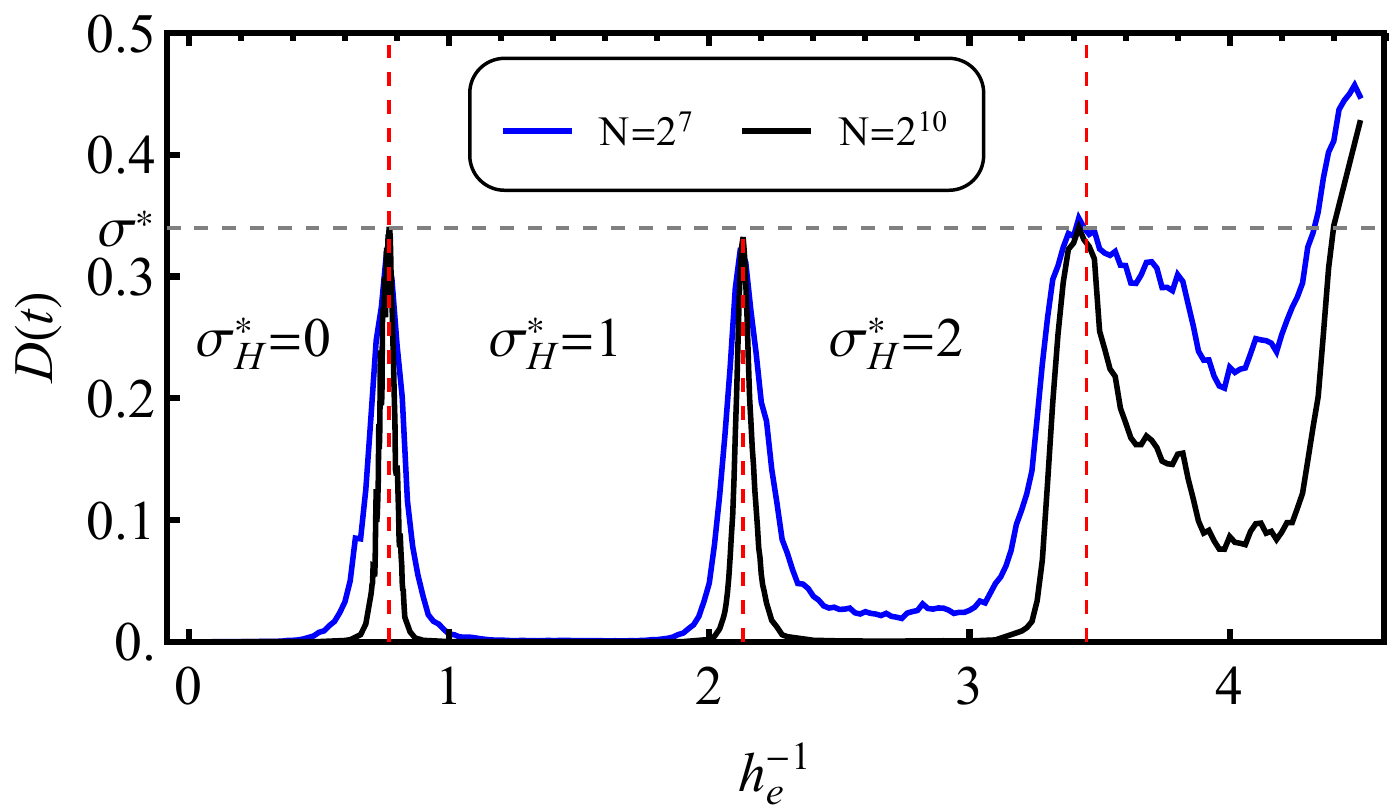}
\caption{Long time diffusion rate of the QKR  model as a function of $h_e^{-1}$ for $N=2^{7}$ (blue line) and $N=2^{10}$ (black line). We take ensemble average over 400 different values of $\alpha \in (0,2\pi)$ and $q \in (0,1)$. We set $t=N^2/4$. The transitions occur near $h_{e}=0.77$, $2.13$, and $3.45$ as indicated with red dashed lines. The diffusion rate converge to a universal value $\sigma^{*}\simeq 0.33$ (gray dashed line) at the these critical points.}
\label{fig:overall}
\end{figure}

The long-time diffusion rate $D(t)$ as a function of Planck's parameter $h_{e}$ for $0\leq h_{e}^{-1}\leq 4.5$ is plotted in Fig. \ref{fig:overall}. The QKR exhibits dynamical localization with $D(t)\rightarrow 0$ as $t\rightarrow \infty$ for a generic $h_{e}$, but undergoes transitions with nonzero diffusion rate near $h_{e}=0.77$, $2.13$, and $3.45$. 

\begin{figure}[tb]
\includegraphics[width=\columnwidth]{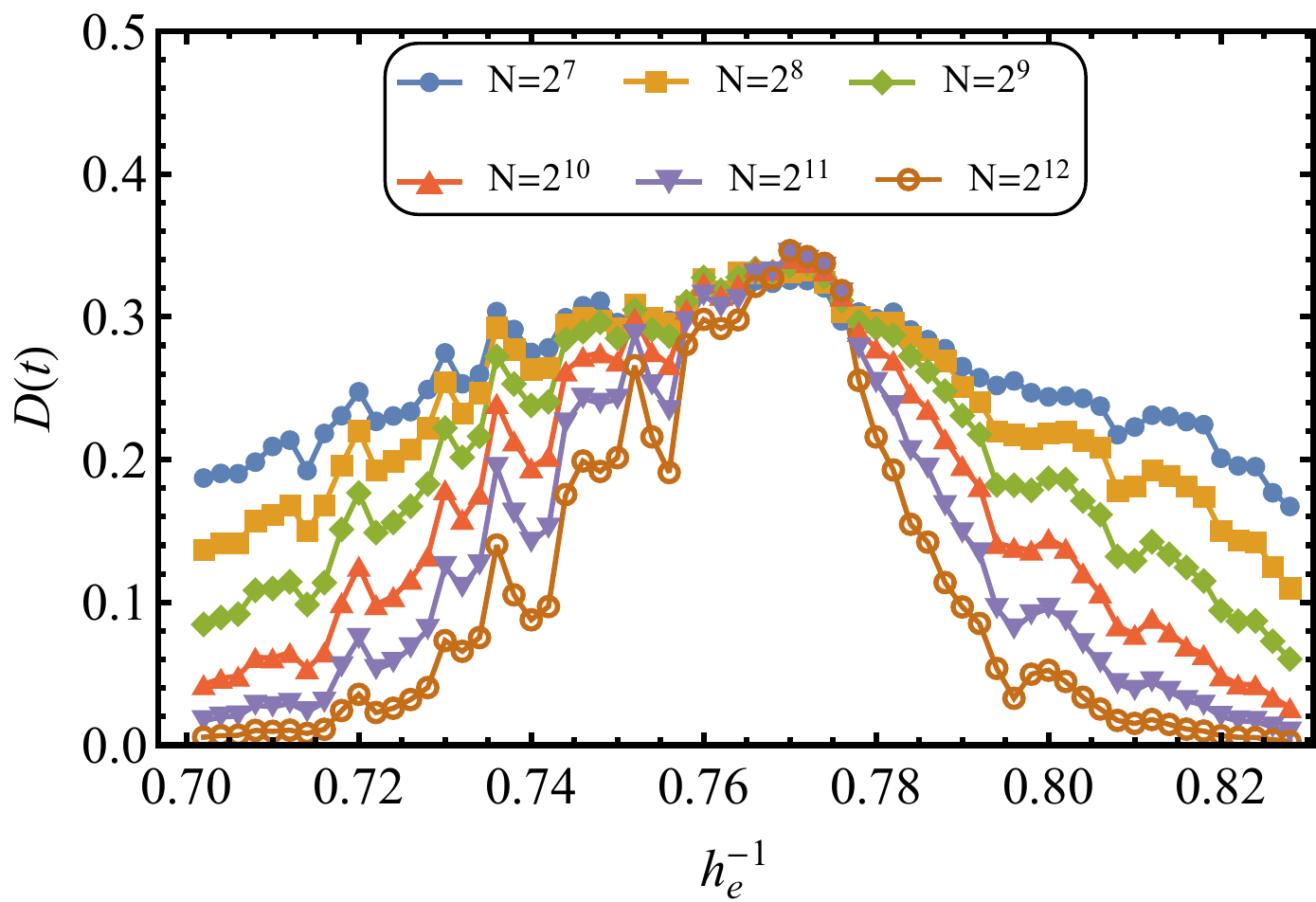}
\caption{Long-time diffusion rate of the QKR model as a function of $h_e^{-1}$ near the critical point $0.77$ for various $N$. We set $t=N^2/4$. }
\label{fig:fig1a}
\end{figure}

\begin{figure}[tb]
\includegraphics[width=\columnwidth]{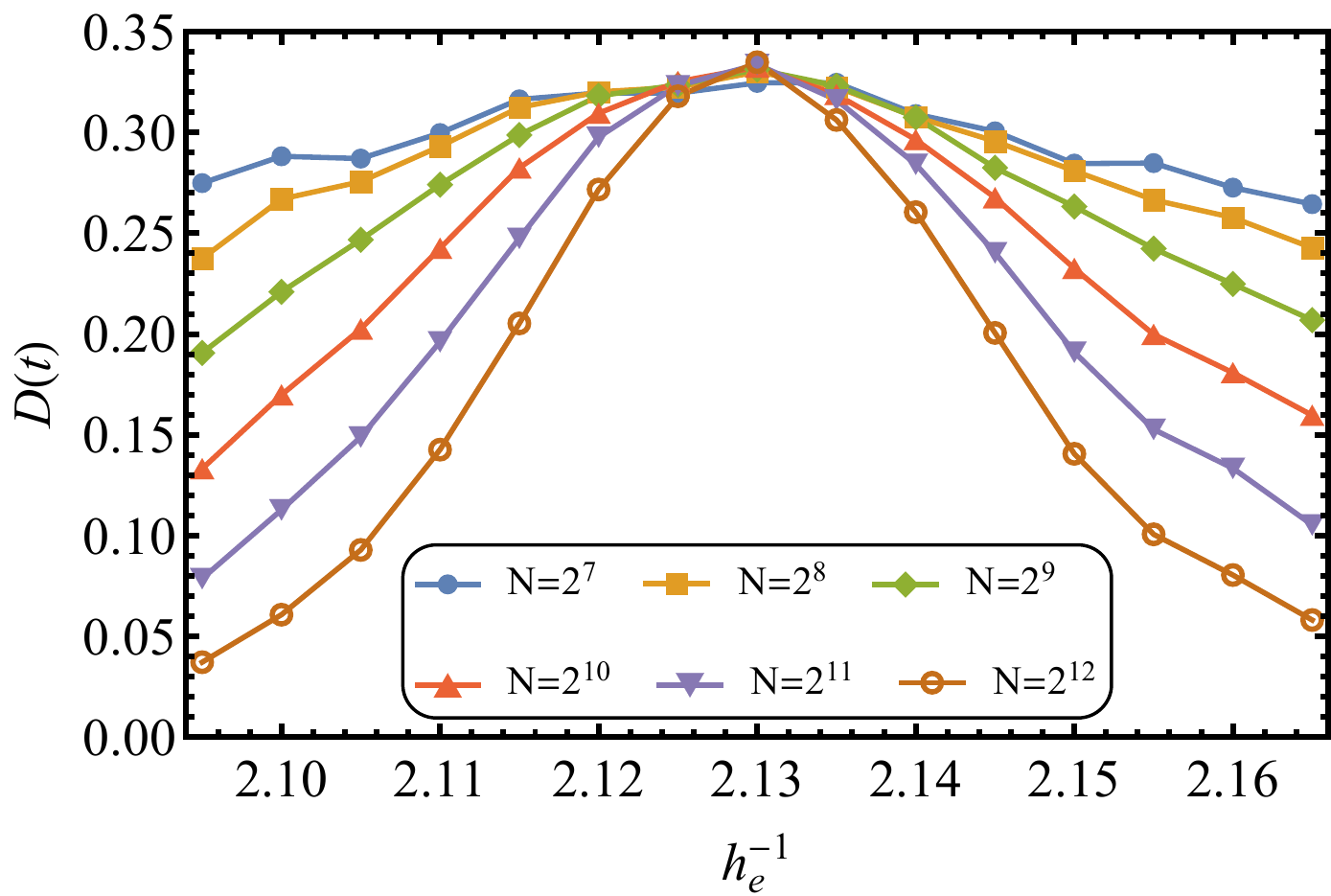}
\caption{Long-time diffusion rate of the QKR model as a function of $h_e^{-1}$ near the critical point $2.13$ for various $N$. We set $t=N^2/4$.}
\label{fig:fig2a}
\end{figure}

\subsection{Finite size scaling analysis}

We then zoom in and carry out extensive simulations near $h_{e}=0.77$ and $2.13$ for various $N$ and set $t=N^{2}/4$. The results are shown in Figs. \ref{fig:fig1a} and \ref{fig:fig2a}. We find that the data near $h_{e}=0.77$ are not quite smooth even after the ensemble average and show a bunch of peaks and dips, which might be attributed to the non-generic behavior induced by rational $h_{e}/4\pi$ \cite{Tian:11, Wang:14}. We thus focus on the data near $h_{e}=2.13$.

In the long-time limit, the diffusion rate $D(t)$ goes to zero for an insulating state with dynamical localization in the momentum space, and approaches a nonzero value for a metallic state. According to the scaling theory of Anderson localization, the diffusion rate obeys the one-parameter scaling law. Near the critical point, the diffusion rate has the scaling form,
\begin{equation}
D(h, t)=\xi^{2-d}F(\xi^{-d}t).
\end{equation}
Here $\xi$ is the localization length, which diverges as $\xi \propto |\delta h|^{-\nu}$ with $\delta h = h_e-h_{e,c}$. $d=2$ is the spatial dimension of the equivalent Anderson model. The truncation of the Hilbert space to $2N$-dimensional introduces a finite lattice size $2N$ in the momentum space, thus the finite-size scaling form is given by
\begin{equation}
D(h,  t, N)=f(t/N^2, hN^{1/\nu})\equiv f(x,y),
\label{eq:scaling}
\end{equation}
with $x=t/N^2$, and $y=hN^{1/\nu}$. $f$ is a non-singular function of its arguments. In the simulations, we choose $t=N^{2}/4$, thus $x=1/4$ is fixed, and expand $f$ into the power series of $y$,
\begin{equation}
f(x_0,y)=\sum^{k_{\mr{max}}}_{k=0}a_{k}y^k.
\end{equation}
The expansion coefficients $a_{k}$'s, the critical point $h_{e,c}$, and the critical exponent $\nu$ are free fitting parameters.

\begin{figure}[tb]
\includegraphics[width=\columnwidth]{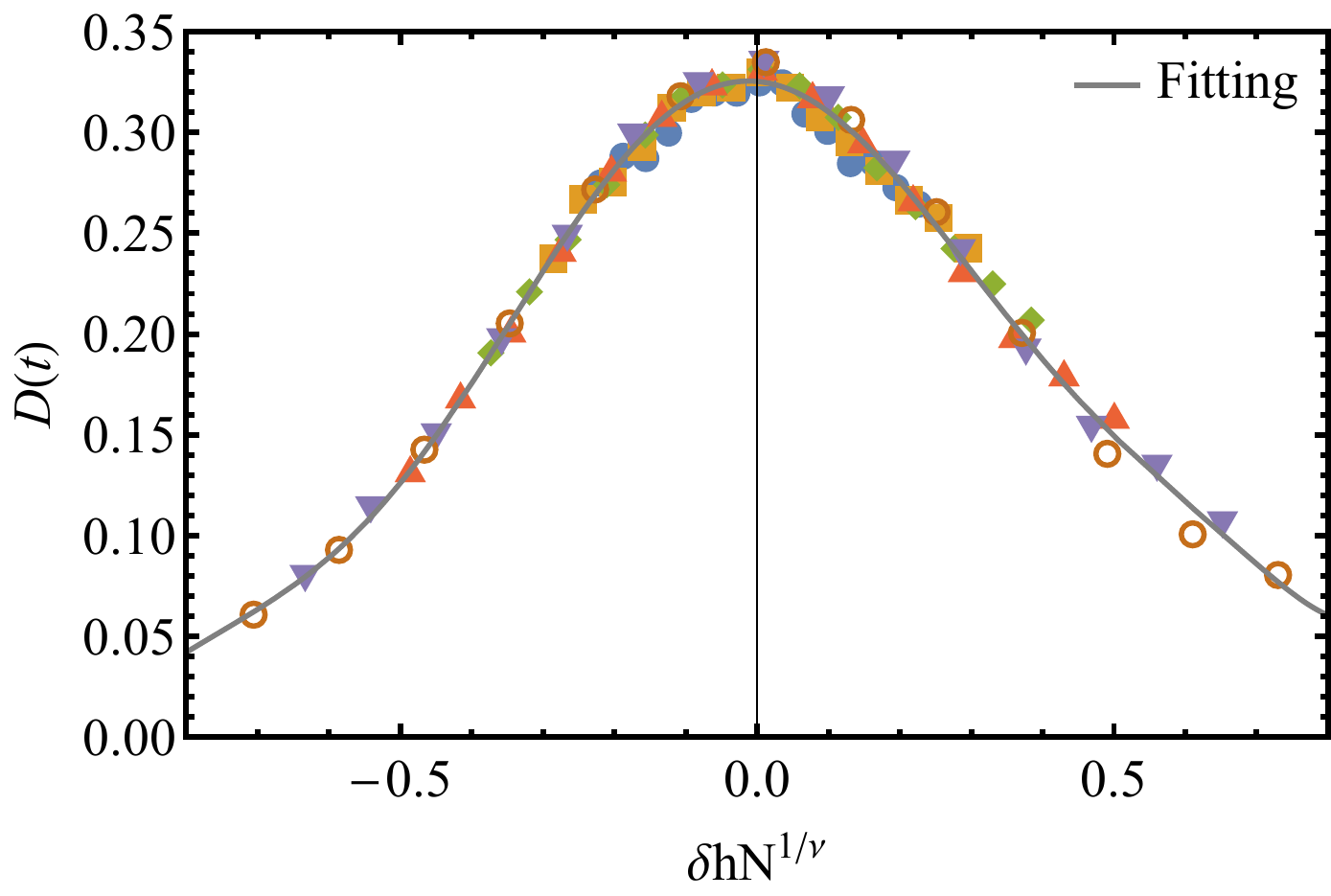}
\caption{Data collapse of the diffusion rate $D(t)$ in Fig. \ref{fig:fig2a} according to finite size scaling form in Eq. (\ref{eq:scaling}).  }
\label{fig:fig1b}
\end{figure}

Applying the above finite-size scaling analysis, we find all data collapse onto a single smooth curve as a function of $\delta hN^{1/\nu}$ (see Fig. \ref{fig:fig1b}). The critical point $h_{e,c}^{-1}=2.1294(3)$, and the critical exponent $\nu=2.62(9)$. This is consistent with that of the IQH plateau transition estimated with the Chalker-Coddington model, $\nu=2.593(5)$ \cite{Chalker:88, Huckestein:95, Slevin:09}, thus we confirm that the critical point of the spin-$1/2$ QKR model belongs to the universality class of the IQH plateau transitions. Moreover, we also give a precise estimate of the universal diffusion rate at the metallic critical point, $\sigma^{*}=0.3253(12)$.

\section{Conclusion}
\label{sec:concl}

To summarize, we have studied the spin-$1/2$ QKR model with extensive numerical simulations. By devising and applying the finite-size scaling analysis near the critical point between different dynamical localization phases, we obtain the numerical estimate of the critical exponent $\nu$ and the universal diffusion rate $\sigma^{*}$ at the critical point. We confirm that the transition belongs to the universality class of the IQH plateau transition.

\begin{acknowledgments}
We would acknowledge helpful discussions with Rui-Zhen Huang. This work is supported by the National Key R\&D Program of China (2018YFA0305800), the National Natural Science Foundation of China (11804337 and 12174387), the Strategic Priority Research Program of CAS (XDB28000000), and the CAS Youth Innovation Promotion Association.
\end{acknowledgments}

\bibliography{qkr}
\end{document}